\documentstyle[12pt,epsfig,psfig,emulateapj]{article}

\lefthead{YADAV J. S.}
\righthead{DISK$-$JET CONNECTION  IN THE MICROQUASAR GRS~1915+105}
\slugcomment{ Accepted for publication in ApJ (Part 1)}

\begin{document}

\title{DISK$-$JET CONNECTION  IN THE MICROQUASAR GRS~1915+105  AND  
IR AND  RADIO EMISSION}

\author{J. S. Yadav{$^1$}}
\affil{Tata Institute of Fundamental Research, Homi Bhabha Road, Mumbai 400 005, India}

\altaffiltext{1}{jsyadav@tifr.res.in}

\begin{abstract}
We  present  evidence of a direct accretion disk $-$ jet connection in the 
Galactic microquasar GRS~1915+105 based on  our analysis of RXTE/PCA data
with a ``spike'' in X-ray light curves. We find that the radio emission
increases as the hardness ratio increases during the low hard state.  
 We suggest that the ``spike'' which
separates the dips with hard and soft spectra marks the beginning of the 
burst phase when the luminosity of the soft X-rays (5$-$15 keV) increases 
by a large factor ($\sim$ 10). This produces a major ejection 
episode  of the synchrotron - emitting plasma termed as ``baby jets'' which 
are associated with infrared (IR) and radio flares of about half an hour period 
widely reported in the literature.
Subsequent short but frequent soft dips produce overlapping faint flares which
result in an enhanced level of quasi-steady emission. 
We discuss the differences between ``baby jets'' and relativistic radio jets
and especially investigate their signatures in X-rays.
\end{abstract}


\section{INTRODUCTION}

Two Galactic X-ray transient sources  GRS~1915+105 and GRO~J1655-40 are known 
to produce relativistic radio jets (\cite{mira:94}; \cite{ting:95}).  The 
combination of relativistic jets and a suspected central black hole 
has earned these 
two objects the name ``microquasars'' as they seem to be stellar mass analogs 
of the massive black hole systems in quasars and other active galactic 
nuclei (AGNs) (\cite{morg:97}; \cite{bell:97b}; 
 \cite{oros:97}).  Since microquasars are much smaller, closer and show faster 
variability than
the extragalactic systems, they are  potential ``laboratories'' 
for the study of black hole accretion/relativistic jet systems.
GRS~1915+105 has shown spectacular X-ray variability since its  
discovery in 1992 (\cite{cast:94}).
Recently, Belloni et al. (2000) have classified a large sample of  RXTE/PCA
observations in 12 separate classes on the basis of their light curves and the
color-color diagrams. Out of these, the $\beta$ class is described as 
the most complex,  
has  all the three basic states of the source (\cite{bell:00}), 
and is always accompanied by
IR/radio flares. The presence of a ``spike'' in the X-ray light curve, which
separates the dips with hard and soft spectra, clearly distinguishes this class
from others. 
The ``spike'' coincides with the beginning of IR flares seen during 
simultaneous X-ray and IR observations, suggesting its role 
in initiating IR flares (\cite{eike:98}). 
These flares are  termed as ``baby jets'' from energy 
consideration.  Another simultaneous observations
of GRS 1915+105 in the X-ray, IR, and radio wavelengths confirm that the IR and radio flares are associated with the X-ray dips (\cite{mira:98}). 

\begin{deluxetable}{llclccc}
\footnotesize 
\tablecaption{SUMMARY OF THE SELECTED OBSERVATIONS OF GRS~1915+105  \label{tbl-1}}
\tablewidth{0pt}
\tablehead{
\colhead{Observation ID} & \colhead{Date} &   \colhead{Exposure} & 
\colhead{PCU\tablenotemark{a}}  &  \colhead{Radio\tablenotemark{b}} & \multicolumn{2}{c}{Infrared (IR) or Radio flares}\\  
\colhead{} & \colhead{} &   \colhead{(s)} &  \colhead{on}  &\colhead{Flux (mJy)}& \colhead{Period (min)}& \colhead{Peak ampl.(mJy)}
}
\startdata
 10408-01-38-00& 1996 Oct 07 & 7000 &5&3& -& -\nl
 20186-03-03-01& 1997 Aug 14 & 10000&5&29& $\sim$ 30& $\sim$ 12 $^\dag$\nl
 20402-01-45-03& 1997 Sep 09 & 10000&4&47& $\sim$ 30& $\sim$ 60 $^\S$\nl
 20402-01-52-01& 1997 Oct 30 & 4000&5&200& $\sim$ 25& $\sim$ 200 $^\ddag$\nl
 20402-01-53-00& 1997 Oct 31 & 10000&5\tablenotemark{c}&170& $\sim$ 30& $\sim$ 150 $^\ddag$\nl
\enddata
\tablenotetext{ } { $^\dag$ IR flares 
(2.2 $\mu$m) \cite{eike:98}, $^\S$ Radio flares (8.3 GHz) \cite{mira:98}, 
$^\ddag$ Radio flares (15 GHz) \cite{fend:99}.}
\tablenotetext{a}{Number of Proportional Counter Units (PCUs) operating at 
the time of observation.} 
\tablenotetext{b}{Radio flux is obtained by  interpolation  of  8.3 GHz 
public domain data from NSF-NRAO NASA except on  1996 Oct. 7 for which 
data are  for 15.2 GHz estimated from figure 4 of Pooley \& Fender (1997).}
\tablenotetext{c}{Except in the time range of 1.20887 $\times 10^8$ to
1.20889 $\times 10^8$ (MJD $\sim$ 50752.6) when only 3 PCUs  were on.}
\end{deluxetable}

At present, the disk-jet connection is indirect at best for relativistic 
radio jets.   Harmon et al. (1997) have shown a long-term correlation
between hard X-ray flux and jet activity in GRS~1915+105. Fender et al.
(1999) have observed four relativistic radio ejections in 1997 
October/November and
have presented high quality MERLIN radio images of ejecta in 400$-$5000 AU
scales which are consistent with ballistic motion. Recently, Eikenberry
et al. (2000) have reported detection of faint IR flares  
and have classified the ejection events associated with 
IR and radio  flares into three classes: (A) the  relativistic events producing
bright superluminal radio jets of $\sim$ 1 Jy with decay times of several 
days (\cite{mira:94}
; \cite{fend:99}), (B) the ``baby jets'' associated with $\sim$ 100$-$200 mJy 
IR and radio flares with decay times of several minutes  
 (\cite{eike:98}; \cite{mira:98}), and (C) faint IR flares with
peak amplitude of $\sim$ 0.5 mJy and duration of 8-10 minutes (\cite{eike:00}).
In  this paper we present the results of our detailed analysis of 
RXTE/PCA observations spread over a year of $\beta$ and $\lambda$ classes
which actually belong to the same class as shown here. 
We  investigate the role of the ``spike'' in initiating the ``baby jets''   
and study the accretion disk corresponding  to IR/radio flares of varying 
peak amplitudes (Table 1). We present  evidence of a direct disk
$-$ jet connection for radio ejections  of class A. We discuss the
dissipation  of accretion energy in the inner part of the disk in terms of 
a  comptonising cloud and  the escaping  mass which produces synchrotron 
radiation. 

\section{OBSERVATIONS AND ANALYSIS}

The observations discussed  here are selected   from the publicly 
available RXTE/PCA data for the  X-ray transient source GRS~1915+105 
(\cite{jaho:96}).   In Table 1 we list details of these observations.
The source was in a high/flaring state during  
these observations.     A 
portion of  2$-$13 keV light curves
for different days added for all available PCA units are shown in Figure 1.
The ``spike'' is present in all the observations shown in 
Figure 1 (all of them belong to   class $\beta$) and IR/Radio 
flares were 
observed during these observations. The peak amplitude of IR/radio flares 
are given in Table 1.
The IR flares
with a spacing of $\sim$30 minutes were observed during simultaneous X-ray/IR 
observation on 1997 August 14 (\cite{eike:98}).  During another simultaneous 
observation in the X-ray, IR, and radio on 1997 September 9,   
 similar flares were seen in both the IR and radio bands (\cite{mira:98}). 
Fender et al.
(1999) have observed four relativistic radio jets during  1997 October/November
 and have seen  radio flares on 1997 October 30-31 in 15 GHz 
band (see their figure 7). The other observation of 1996 October 7 belongs to  
class $\lambda$ (shown in Figure 2) which   
is  a ``radio-quiet'' state. Muno et al. (1999) have defined the 
``radio-quiet'' state when the radio flux in the 8.3 GHz band (or the 
15 GHz band) 
is less than 15 mJy.  The general approach of our analysis is described
elsewhere (\cite{yada:00}).  The two 
X-ray colors HR$_1$ and HR$_2$ are defined as the ratio of the flux in 
the 5$-$13 keV band to flux in the 2$-$5 keV band and the ratio of the flux in 
the 13$-$60 keV band to flux in the 2$-$13 keV band respectively. The sub-second
rms variability is calculated from 0.1 s data. 

To study the spectral properties during these observations, we have used
simultaneous fits to the PCA data in the energy range of 5$-$60 keV and 
the HEXTE data 
in the energy range of 15$-$100 keV.
We used ``Standard 2'' mode   PCA data  (16 s time resolution)  
with  128 channel spectra  and  1\% systematic error added.
The 64 channel archive mode data of HEXTE (cluster 0) was rebinned to improve the statistics of the 
counts.  We have used a sum of a 
multicolor disk black body  (DISKBB) and a power law  to approximate 
the energy spectrum (\cite{yada:00b}).

\section{RESULTS AND INTERPRETATION}

The light curves  seen on 1997 October 31 (class $\beta$) and on 1996 
October 7 (class $\lambda$) are shown in the top panels of Figure 2.
The HR$_2$ color is shown in the middle panels and   
sub-second mean variability is shown in the bottom panels of Figure 2. 
In the top panels, durations of the low hard state (quiescent phase), the high 
soft state (burst phase) and  the soft dips (low soft state) are  marked 
by `C', `B' and `A', respectively.  Although Belloni et al. (2000) 
have found short soft dips of state `A' in the class 
$\lambda$ (in the beginning of  burst phase and during the fast fluctuations),
we have marked it whole as state `B'. This is done to emphasize the long  
dips of state `A' found in class $\beta$. The HR$_2$ is high (0.09$-$0.12)  
during the quiescent phase (state C) whereas it is low (0.03$-$0.06) during 
the burst phase (state B) and the soft dips (state A). 
The similarity of 
the two plots in the middle panels suggests that 
the ``spike'' in the light curves of class $\beta$ marks the beginning of the 
burst phase. This  is shown by a vertical arrow in the top left panel. The mean 
variability plotted in bottom panels is consistent with 
this suggestion. The mean variability is 2$-$5 \% during the burst phase, 
10$-$15 \% during the quiescent phase and 5$-$9 \% during the soft dips.
In  the region of fast fluctuations 
during  phase B of both the classes, the mean 
variability is mostly in the range of 5$-9$\% indicating the presence of
short soft dips during these fast oscillations.
These results suggest that classes $\lambda$ and $\beta$ of Belloni et al.
(2000) are very similar
except for (1) the presence of long soft dip in the beginning of the burst 
phase   
in case of class $\beta$ while it is short and  shallow in case of 
class $\lambda$ (see figure 4 of Belloni et al. (2000)), 
and (2) the short but frequent soft dips during the
fast oscillations  are relatively 
deeper and for longer times in  case of class $\beta$ than that in 
class $\lambda$.  It may be noted here that the HR$_2$ distinguishes the quiescent phase from
other two phases (the burst phase and the soft dips) while the mean 
variability highlights the difference between the burst phase and the
soft dips.

\begin{deluxetable}{lccccc}
\footnotesize 
\tablecaption{SIMULTANEOUS SPECTRAL FITS TO PCA AND HEXTE DATA IN THE ENERGY
RANGE 5$-$100 keV}
\tablewidth{0pt}
\tablehead{
\colhead{Date} &   \colhead{Quiescent phase} & \colhead{Soft dip} &\colhead{Burst phase} &  \colhead{Luminosity ratio} & \colhead{Luminosity ratio}\\ 
\colhead{} &   \colhead{(hard dips)(C)} &  \colhead{(A)} & \colhead{(B)} &\colhead{(5$-$15 keV)}& \colhead{(20$-$100 keV)}
}
\startdata
1996 Oct 07 & T$_{in}$= 1.11$\pm$0.03&&T$_{in}$= 2.23$\pm$0.01&B/C=10.16&B/C=0.52\nl
& $\Gamma$= 2.23$\pm$0.03&&$\Gamma$= 3.59$\pm$0.04&&\nl
& $\chi^2_{\nu}$= 0.75&&$\chi^2_{\nu}$= 1.10&&\nl
&R$_{in}$=37.9  &&R$_{in}$=25.2 &&\nl
&$\dot{m}$=2.19 $\times 10^{18}$&&$\dot{m}$=1.08 $\times 10^{19}$&&\nl
\hline
1997 Oct 30 & T$_{in}$= 0.87$\pm$0.26&T$_{in}$= 2.01$\pm$0.04&T$_{in}$= 2.13$\pm$0.02&B/C=3.01&B/C=0.43\nl
& $\Gamma$= 2.78$\pm$0.02&$\Gamma$= 3.77$\pm$0.04&$\Gamma$= 3.57$\pm$0.04&A/C=1.46&A/C=0.20\nl
& $\chi^2_{\nu}$= 1.40&$\chi^2_{\nu}$= 0.83&$\chi^2_{\nu}$= 0.80&B/A=2.06&B/A=2.17\nl
&R$_{in}$=44.7  &R$_{in}$=16.2  &R$_{in}$=23.2 &&\nl
&$\dot{m}$=1.40 $\times 10^{18}$&$\dot{m}$=1.89 $\times 10^{18}$&$\dot{m}$=6.94 $\times 10^{18}$&&\nl
\enddata
\tablenotetext{}{T$_{in}$ is the temperature of the inner accretion disk in
keV, $\Gamma$ is the power law index,  R$_{in}$ is the characteristic radius
of the inner accretion disk in km with error of $\le$ 10\% and $\dot{m}$ is
the accretion rate in g s$^{-1}$.}
\end{deluxetable}

The class $\beta$ has been  studied simultaneously in IR and radio bands. 
Eikenberry et al. (1998) have compared X-ray and IR profiles during
1997 August 14 observations and have found that the spikes in the X-ray band  
coincide with the beginning of IR flares (see their figure 3). The IR 
intensity reaches 
its peak flux shortly  after the X-ray  peak. The IR
flares have decaying phases very similar in their smoothness and time scale
to the rising phases. On the other hand, 
the X-ray  flux begins the fast oscillations  after reaching their peak 
level as mentioned earlier. 
An excess in IR is seen during the fast X-ray oscillations. This excess
in IR is explained in terms of many faint IR flares superposed on one 
another (\cite{eike:00}). If we assume that each X-ray oscillation has an
associated faint IR flare, an IR excess of 1.3 mJy is estimated,  which is 
consistent with the observed excess of $\sim$ 1.0 mJy for 1997 August 14
observations. During another simultaneous observation of $\beta$ class in
IR and radio on 1997 September 9, Mirabel et al. (1998)  confirmed 
the role of the spike in
initiating IR flares and have found that the radio flares follow the IR 
flares with a time delay consistent with broad band synchrotron emission. 
The profiles of IR and radio flares are
quite similar. In contrast, the $\lambda$ class belongs to radio-quiet state. 
Pooley \& Fender (1997) have observed  GRS~1915+105 in radio   
during 1996 October but no radio flares were reported during 
observations of $\lambda$ class X-ray activity  on  1996 October 7. 
 
To investigate further the similarities and differences  between $\beta$ 
and $\lambda$ classes, 
we have studied the mean X-ray color HR$_1$, which is a  measure of  
soft photons and hence the disk temperature. On the other hand, 
HR$_2$ is a measure 
of hard photons and a indicator of appearance/disappearance of the 
advective/halo disk (\cite{yada:99}).
The mean HR$_1$ during the quiescent phase is plotted as a function of mean
quiescent flux in Figure 3 for all the observations in Table 1.   
 The mean HR$_1$ increases with the quiescent flux 
and follows a single path. 
The  HR$_2$ during the quiescent phase also shows similar results except 
it decreases with the quiescent flux (in the inset of Figure 3).  These results 
reinforce the above suggestion that classes $\beta$ and $\lambda$ are very 
similar and actually belong to a single class.    

We have studied the spectral properties of both $\beta$ and $\lambda$ classes 
using simultaneous fits to PCA and HEXTE
data in the energy range 5$-$100 keV as described in the previous section. 
The segments of the quiescent phase, the burst phase, and the soft dips are analysed 
separately. We have selected  $\beta$ class data of 
1997 October 30  (Figure 1c) for spectral analysis as the first 
relativistic jet observed by Fender et al. (1999)
was close to this X-ray observation, and also the $\lambda$ class  data of 
1996 October 7 (Figure 2 right panels). The characteristic  radius of the inner
disk, R$_{in}$ is derived from the best-fit spectral parameters and the known 
distance and inclination (the latter is taken as the same as the inclination of 
the radio jets to the line of sight). From R$_{in}$ and T$_{in}$, we infer
the mass accretion rate using $ \dot m = 8 \pi R_{in}^{3} \sigma T_{in}^{4} /
3GM $ where $\sigma$ is Boltzmann constant, G is the gravitational constant 
and M is the mass of the black hole. The  value of R$_{in}$, which is smaller
during the
burst phase, cannot be less than  the innermost stable orbit around a 
black hole. For a Schwarzschild black hole ($R_{in} \ge 6GM/c^{2}$), 
an R$_{in}$ of $\sim$ 20 km during the burst phase puts M $\lesssim$ 
2.4 M$_{\odot}$
(\cite{bell:97a}). The best-fit parameters are given in Table 2. 
 Note that the derived values for the inner disk radius can be 
an underestimate due to scattering 
effects and  the approximations  made in fitting 
the Comptonised part of the spectrum as a power-law (see Shrader \& 
Titarchuk 1998).  We use these  numbers here  only for a 
qualitative description.

The results in Table 2 are consistent  with the general picture: the spectrum is
hard during the quiescent phase ($\Gamma \sim$ 2.5) and it is soft during 
the burst phase ($\Gamma \sim$ 3.6). The soft X-ray  luminosity (5$-$15 keV)
increases over a factor of  ten  during the burst phase while the hard X-ray 
 luminosity (20$-$100 keV) reduced to half than  that during  the quiescent 
phase for 1996 October 7 observations. The total X-ray luminosity (5$-$100 
keV) drops by a factor of $\sim$ 5.5 during the quiescent phase than  that 
in the burst phase. Yadav et al. (1999) have suggested that the change of 
states during the flaring of the source (like during class $\lambda$) is due to 
the appearance/disappearance of an advective/halo disk. If we assume the above
scenario, most of the energy ($\sim$ 80\%) disappears during the 
quiescent phase through advection into black hole (\cite{nara:97}). 
During the quiescent phase,
 the total X-ray flux  in the 5$-$100 keV band is increased from a value of 
7.1 $\times$ 10$^{-9}$ erg 
s$^{-1}$ cm$^{-2}$  on 1996 October 7 to a value of 16.4 $\times$ 10$^{-9}$ 
erg s$^{-1}$ 
cm$^{-2}$ on 1997 October 30 while the B/C ratio for  5$-$15 keV X-ray is 
reduced from  10.16 to a value of 3.01 (see Table 2). The simplest 
explanation of these results may be 
that the disk disrupts due to mass ejection on 1997 October 30 before X-ray
luminosity could reach its peak value during the burst phase.
 During the burst
phase the 5$-$100 keV X-ray flux is $\sim$ 3.4 $\times$ 10$^{-8}$ erg 
s$^{-1}$ cm$^{-2}$ on 1997 October 30 whereas  it is $\sim$ 3.9 $\times$ 
10$^{-8}$ erg s$^{-1}$ cm$^{-2}$ on 1996 October 7 when short, soft dips 
were observed by Belloni et al. (2000) as discussed earlier which 
suggest the beginning of mass ejection (radio flux is still low,
see Table 1). These results indicate  a luminosity 
threshold to start the disk evacuation which produces the long,  
soft dip after a spike in the X-ray light curve.
 
The B/A luminosity ratio is $\sim$ 2 in the whole range of 5$-$100 keV on 
1997 October 30 (see Table 2) while  $\Gamma$ increases to 3.77 during 
the soft
dip from a value of 3.57 during the burst phase suggesting a 
portion of the inner disk and the halo are blown away/disappears 
during the soft dips. Feroci et al. (1999) have also proposed of a  
 disappearance of the inner disk during soft dips 
from  BeppoSAX observations of GRS~1915+105. Note that the viscous time 
scales of the 
inner disk and that of the halo have almost same value of 
about 1 s which agrees with the fall time scale of the soft dips 
(\cite{bell:00}; \cite{yada:99}). As the halo disappears more disk is  
observable reducing the R$_{in}$ during the soft dips (see Table 2). 
Note here that the ``spike'' marks the beginning of the burst phase as
described earlier. The accretion rate is 
reduced from $\sim$ 6.94 $\times$ 10$^{18}$ g s$^{-1}$ during the burst 
phase to  $\sim$ 1.89 $\times$ 10$^{18}$ g s$^{-1}$ during the soft dip 
phase  
suggesting $\sim$ 5 $\times$ 10$^{18}$ g s$^{-1}$ mass is being blown away
during these soft dips. In the above scenario, the missing energy of
X-rays in the 5$-$100 keV range during the major soft dips is $\sim$ 
1.4 $\times$ 10$^{39}$ erg s$^{-1}$ using the known distance of the source.
 Fender et al. (1999) have estimated a minimum power of $\sim$ 
2 $\times$ 10$^{39}$ erg s$^{-1}$ is required  for  12 hours of  
rising time of the relativistic radio jets observed 
on 1997 October/November on the basis of 
equipartition of energy. Including
one proton for each electron approximately doubles this energy and 
requires  a mass flow rate of $\ge$ 10$^{18}$ g s$^{-1}$. 
The energy requirement is broadly in agreement with our estimate of 
missing  X-ray energy (5$-$100 keV)  of  $\sim$ 1.4 $\times$ 10$^{39}$ 
erg s$^{-1}$  as X-rays below 5 keV should significantly
increase this estimate. During the soft dip, the accretion rate drops by
$\sim$ 5 $\times$ 10$^{18}$ g s$^{-1}$ from  that  during the burst phase
which is in agreement with the required mass outflow rate.

The average radio flux is given in Table 1 for all X-ray observations 
discussed here.  As HR$_1$ increases the radio flux also increases.  
The peak amplitude of IR/radio flares also increases with  HR$_1$  
during class $\beta$ observations (Table 1).
Our spectral analysis  has suggested that X-ray luminosity increases
with HR$_1$.
In  Figure 4 we show the  radio flux at 2.25 GHz from the NSF-NRAO-NASA 
Green Bank Interferometer  for 1997 October 30-31 (data for $\sim$ 48 hours)
on the y-axis (left). The X-ray average color HR$_1$ 
during the quiescent phase 
is plotted on the y-axis (right) during the same time 
(data for $\sim$ 24 hours).  
 The   timing of 
our X-ray data falls between the first and second relativistic radio jets
observed by   Fender et al. (1999).     
Clearly,  results shown in Figure 4
suggest a correlation between the radio and X-ray data. A least-square fit
to the HR$_1$  and  radio data results a relation $radio_{Jy}
= 0.615 \times HR_{1} - 0.322$ with correlation coefficient close to one.
This linear relation is based on a limited set of data and is valid in 
a narrow range of HR$_1$ from 0.94 to 1.2 during the relativistic radio
jets. The issues like extrapolation of this relation beyond the valid
range of HR$_1$, its applicability during the baby jets and fitting a 
non-linear relation (instead of a  linear relation) will be 
discussed separately (\cite{yada:00b}).
In Figure 5, we show the decay profile of  this  radio flare  
observed during 1997 October 30 - November 6
 (offset by 50748.2 MJD).  The profile is 
consistent with  an 
exponential decay having a time constant of 3.157 day (dotted line). 
We have also plotted  the decay profile of another 
radio flare of class A observed on
1999 June 8-15 (offset by 51334.5 MJD) in Figure 5. This flare has been 
observed in the X-ray band by IXAE/PPC (\cite{naik:00}) as well as 
partially by RXTE/PCA. The decay profiles 
of these two flares of class A ($\sim$ two  years apart) are  identical except
some deviation in the middle (on days 6 through 8).  
The average X-ray color HR$_1$ 
during the low hard state is also plotted in Figure 5  
for both the radio flares using the linear relation between the HR$_1$ and 
 radio flux density discribed above. These points are
in good agreement with the decay profiles over the  duration of observations 
when the radio flux drops by a factor of $\sim$ 2. 

Let us summarise the physical picture  emerging here. When the source
changes state from a quiescent phase to the burst phase, the soft X-ray 
luminosity increases by a large factor ($\sim$ 10). If the total X-ray 
luminosity
is above a certain threshold, it promptly triggers a major ejection event of 
synchrotron - emitting  plasma producing a large  soft dip  after a spike
in the X-ray light curve.   Subsequent short but frequent dips and the 
undetectable continuous ejection
during the burst phase, and probably also during the quiescent phase 
(if HR$_1$
is high), produce an enhanced level of quasi - steady emission.  
This  conclusion agrees 
with the  observations of faint IR flares  during the hard 
state which start even before the X-ray oscillations (\cite{eike:00}).
On 1997 October 30, the total 5$-$100 keV X-ray flux  during the 
quiescent phase was  similar   to that  found during a  ``plateau'' state 
 ($\sim$ 1.7 $\times$ 10$^{-8}$ erg s$^{-1}$ cm$^{-2}$) with the enhanced  
quasi-steady radio emission of $\sim$ 50 mJy (15 GHz) lasting for 20 days 
(\cite{fend:99}) which would suggest mass ejection during the quiescent phase.
  It is 
also consistent with the value of k = 2 (suggesting continuous jets) 
derived from
the flux densities of approaching and receding  components of the relativistic
radio jets  observed on 1997 October/November (\cite{fend:99}).

  The radio
emission during ``baby jets'' is consistent with synchrotron emission from 
an adiabatically expanding small cloud  with a flat radio spectrum 
(\cite{fend:97}; \cite{eike:98}). A magnetic field of 8-16 G and  total
energy $\ge$ 10$^{40}$ erg for each half hour flare  are estimated assuming 
 equipartition. The dominant decay mechanism is  adiabatic expansion 
losses.
For a pure adiabatic expansion, the synchrotron radio luminosity 
I$_{\nu}$(r) $\propto$ r$^{-2p}$, where r is the distance 
(eq. (19.34) of Longair
(1994)). This changes to I$_{\nu}$(r) $\propto$ r$^{-2}$ for a flat radio
spectrum. For a cloud of size $\sim$ 10 AU (minor axis at 15 GHz) and a 
relativistic speed of  $\sim$ 10 AU hr$^{-1}$ (\cite{dhaw:00}), a unit 
peak radio luminosity will decay to a value of $\sim$ 0.44 in 15 minutes time
which is in good agreement with the observations (\cite{eike:98}; 
\cite{mira:98}; \cite{pool:97}). 

   The radio emission during the relativistic  jets is consistent with 
synchrotron emission from a extended radio cloud  with ballistic 
motion on the scale of 300-5000 AU and steepening of the  spectrum 
with time ($\alpha$ changes from 0.5 to 1.0 in a few days). Atoyan \&
Aharonian (1999) have shown that a single population of relativistic
particles accelerated at the time of ejection cannot explain the radio
emission from the relativistic  jets and it requires continuous
replenishment of energetic particles with  energy dependent losses. 
Recently, Kaiser et al. (2000) have developed an internal shock model for 
the origin of relativistic radio jets in  microquasars assuming
quasi-continuous jet ejection. In this model, much of  the energy needed to 
produce the radio emission is  stored in the material of the continuous jet
which was ejected by the central source prior to the formation of the shock
fronts. The shock is caused by the collision  of shells of jet material 
moving at different velocities. The formation of the internal shock `lights up'
the relativistic jets by accelerating  particles which emit the 
observed synchrotron radiation. The relativistic jet mode in   which  
 strong internal shocks are produced,   
is thought to have strong variations in the jet speed. 
This is consistent with the observed X-ray color HR$_1$ $\geq$
1.1 in the initial phase  of both the relativistic radio jets (Figure 5),  
which drops 
to a value of 0.94 in less than a day's time (Figure 4). On the other hand,
``baby jets'' are  supposed to represent a relatively stable mode where weak
internal shocks (if any) do not play any significant role in the  
synchrotron emission.
This is in agreement with observed values of X-ray color HR$_1$ which remains
in the range of 0.9$-$1.0 for both the data sets of 1997 August 14 and 
September 9 (Figure 3).
 
\section{Conclusions}
We  present    evidence
of a direct accretion disk $-$ jet connection for relativistic radio jets.
We have provided here an explanation for the most complex
light curves with a ``spike''  observed  to date in
the Galactic microquasar GRS~1915+105. 
We suggest that the ``spike'' which
separates the dips with hard and soft spectra marks the beginning of the 
burst phase, when the luminosity of the soft X-rays  increases 
by a large factor ($\sim$ 10). This produces a major ejection 
episode of the synchrotron - emitting plasma termed as ``baby jets''  with 
the half-hour spacing  widely reported in the literature.
Subsequent short but frequent soft dips produce overlapping faint flares which
result in an enhanced level of quasi-steady emission.  The radio
emission during ``baby jets'' is consistent with synchrotron emission from 
an adiabatically expanding small cloud (within a distance of few tens of AU of the inner 
accretion
disk) with a flat radio spectrum  and a decay time of 
several minutes. On the other hand, the  radio emission during the 
relativistic radio jets is consistent with synchrotron emission from the 
extended  
radio cloud  with ballistic motion on a scale of a few hundreds to few 
thousands AU and a decay time of a few days. These jets require continuous
replenishment of energetic particles. Our results support  the relativistic
jet model with  quasi-continuous mass  ejection and strong variations in the 
jet speeds required for the formation of 
 strong internal shocks to  continuously accelerate the particles. 

\acknowledgments
Author thanks the referee for his very useful comments which has greatly
improved the contents and the presentation of the paper.
He thanks A. R. Rao and K. P. Singh for their valuable comments on the 
manuscript. He also thanks  RXTE  and NSF-NRAO-NASA Green Bank Interferometer 
teams for making 
their data publicly available.  The Green Bank Interferometer is a facility 
of the National Science Foundation operated by the NRAO in support of NASA 
High Energy Astrophysics programs. 


\clearpage

\begin{figure}
\psfig{file=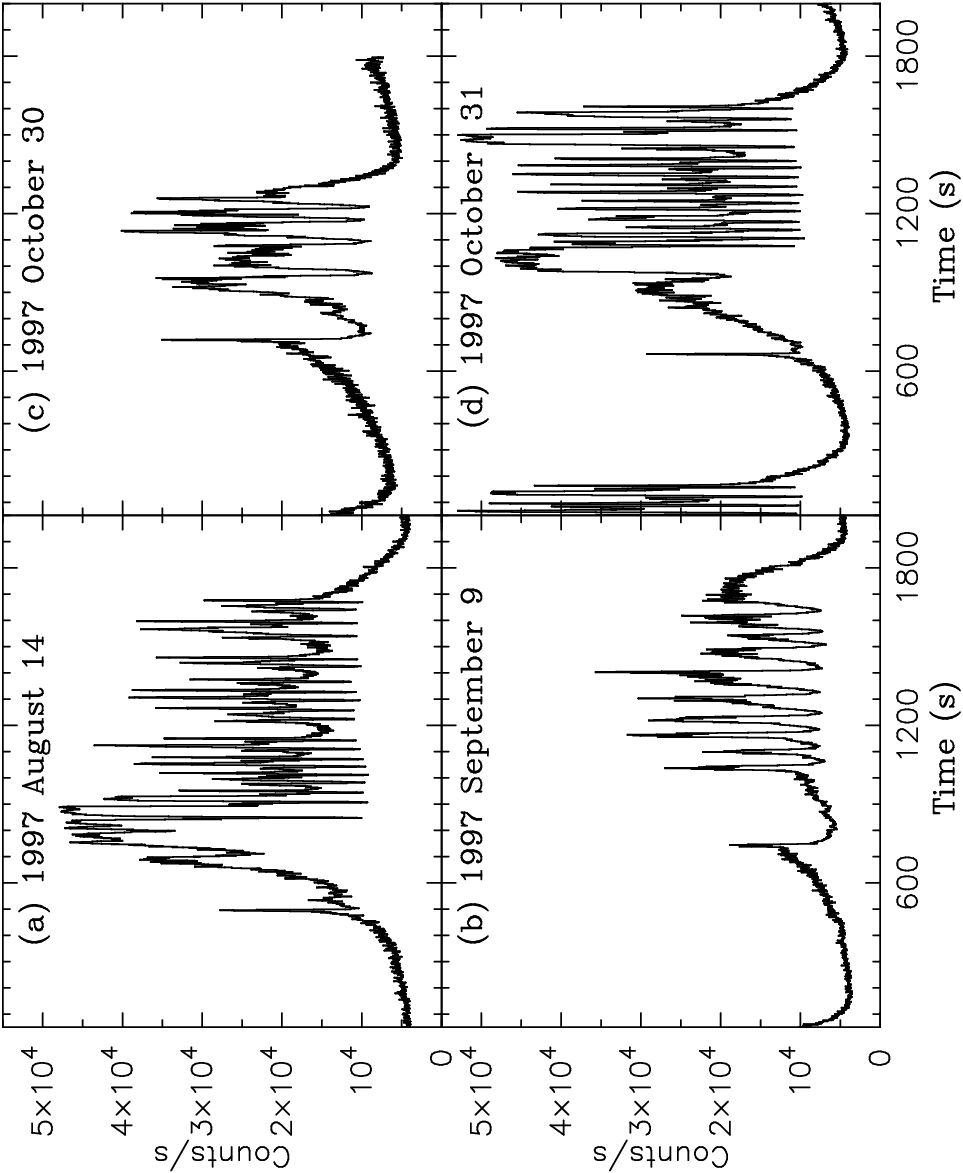}
\figcaption[fig1.eps]{The 2$-$13 keV light curves of GRS~1915+105 observed on
(a) 1997 August 14, (b) 1997 September 9, (c) 1997 October 30, and (d) 1997
October 31. The presence of a ``spike'' clearly shows that all these belong to
class $\beta$. Only 4 PCUs were operating  on  1997 September 9.  \label{fig1}}
\end{figure}

\clearpage

\begin{figure}
\psfig{file=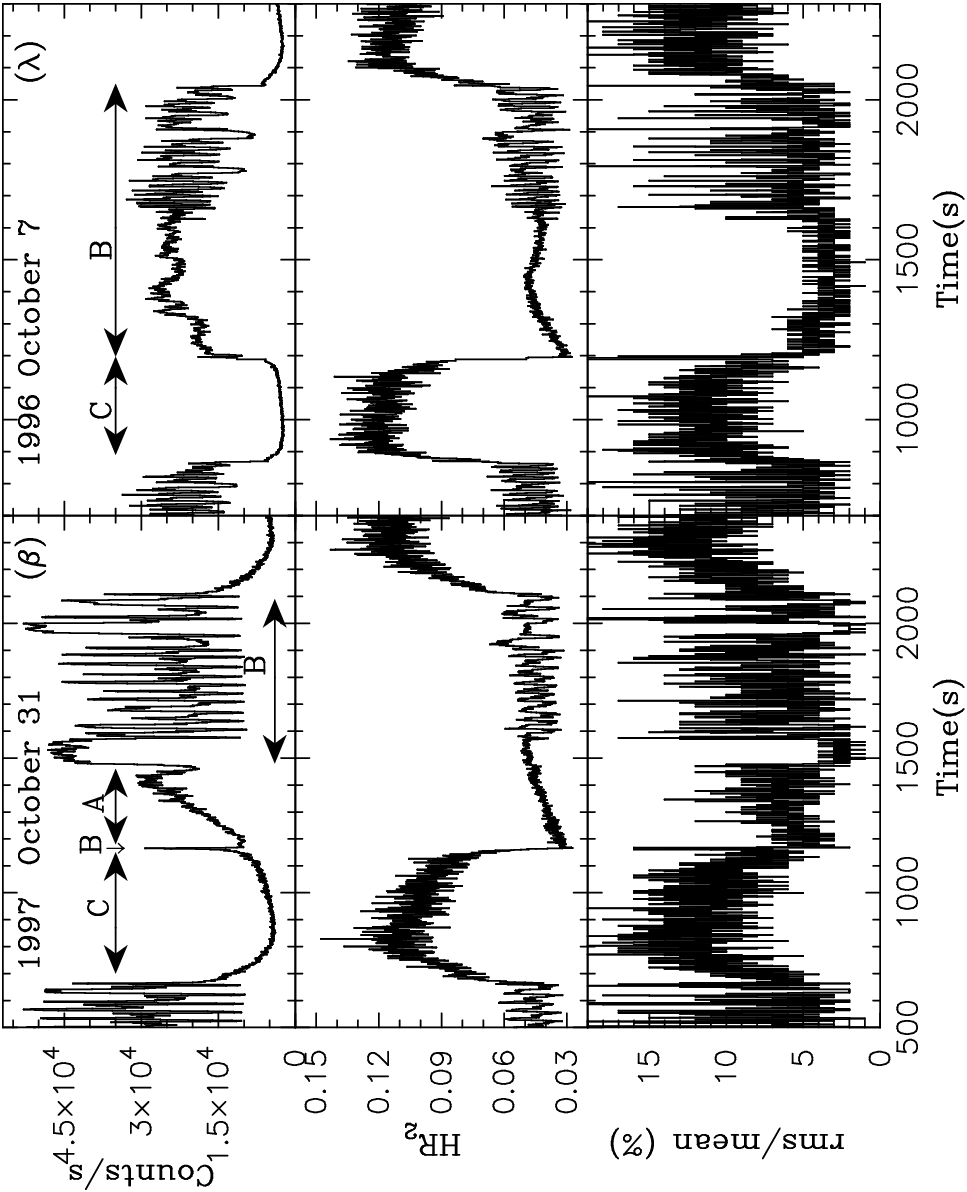}
\figcaption[fig2.ps]{The 2$-$13 keV light curves of GRS~1915+105 observed on
1997 October 31 (class $\beta$) and 1996 October 7 (class $\lambda$) are 
shown in the top panels. The color HR$_2$ is shown in the middle panels and  
sub-second mean variability is shown in the bottom panels. In the top panels, 
durations  of the low hard state (quiescent phase), the  high soft state 
(burst phase) and  the soft dips (low soft state) are marked by `C', `B' 
and `A', respectively (for details see the text).  \label{fig2}}

\end{figure}

\clearpage
\begin{figure}
\psfig{file=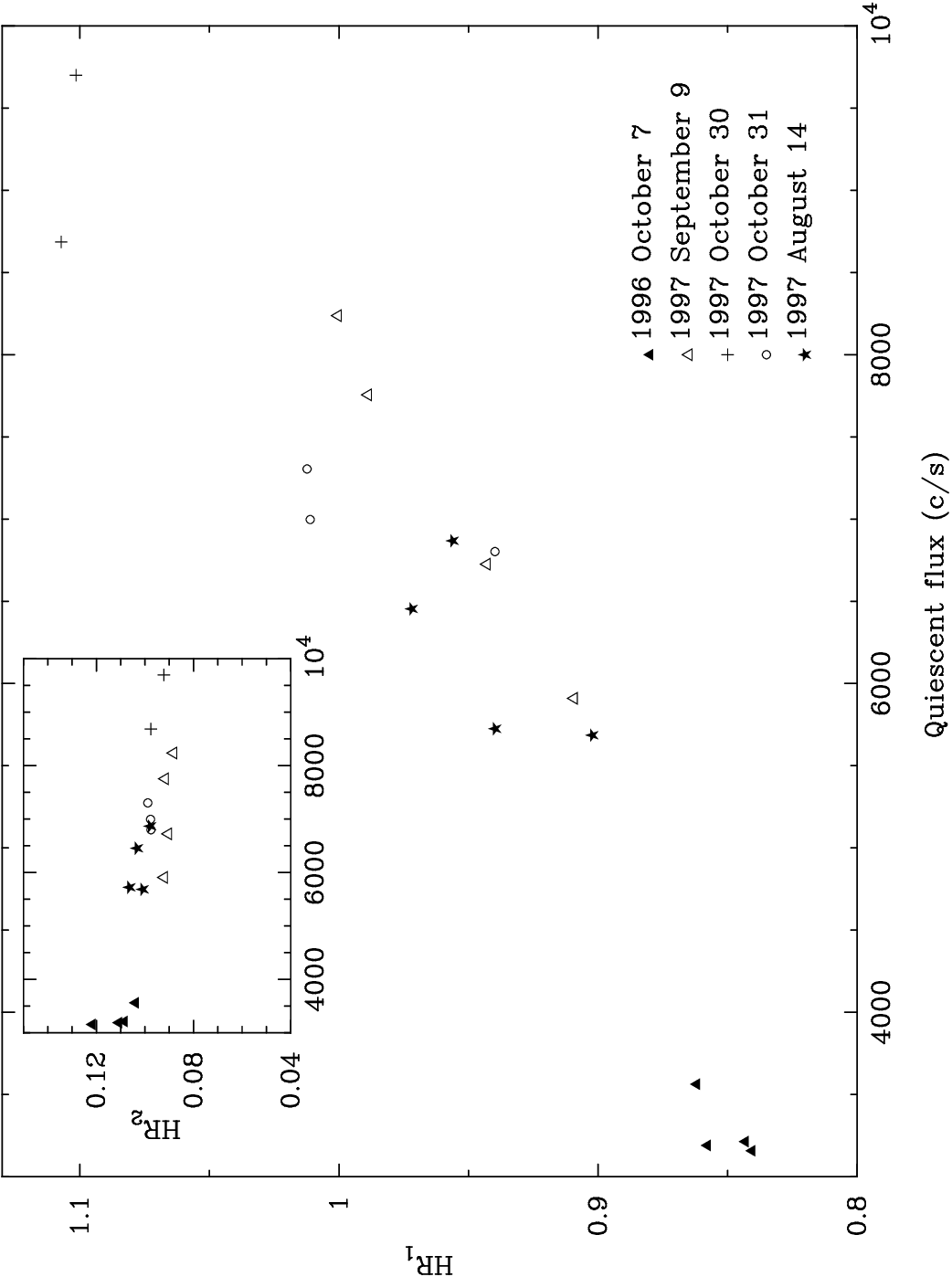}
\figcaption[fig3.ps]{The average HR$_1$ during the quiescent phase (state C) 
is 
plotted as a function of average quiescent flux. The average HR$_2$ during 
the quiescent phase is shown in the inset. The HR$_1$ and HR$_2$ are defined as the ratio of the flux in 
the 5$-$13 keV band to flux in the 2$-$5 keV band and the ratio of the flux in 
the 13$-$60 keV band to flux in the 2$-$13 keV band, respectively. \label{fig3}}

\end{figure}

\clearpage

\begin{figure}
\psfig{file=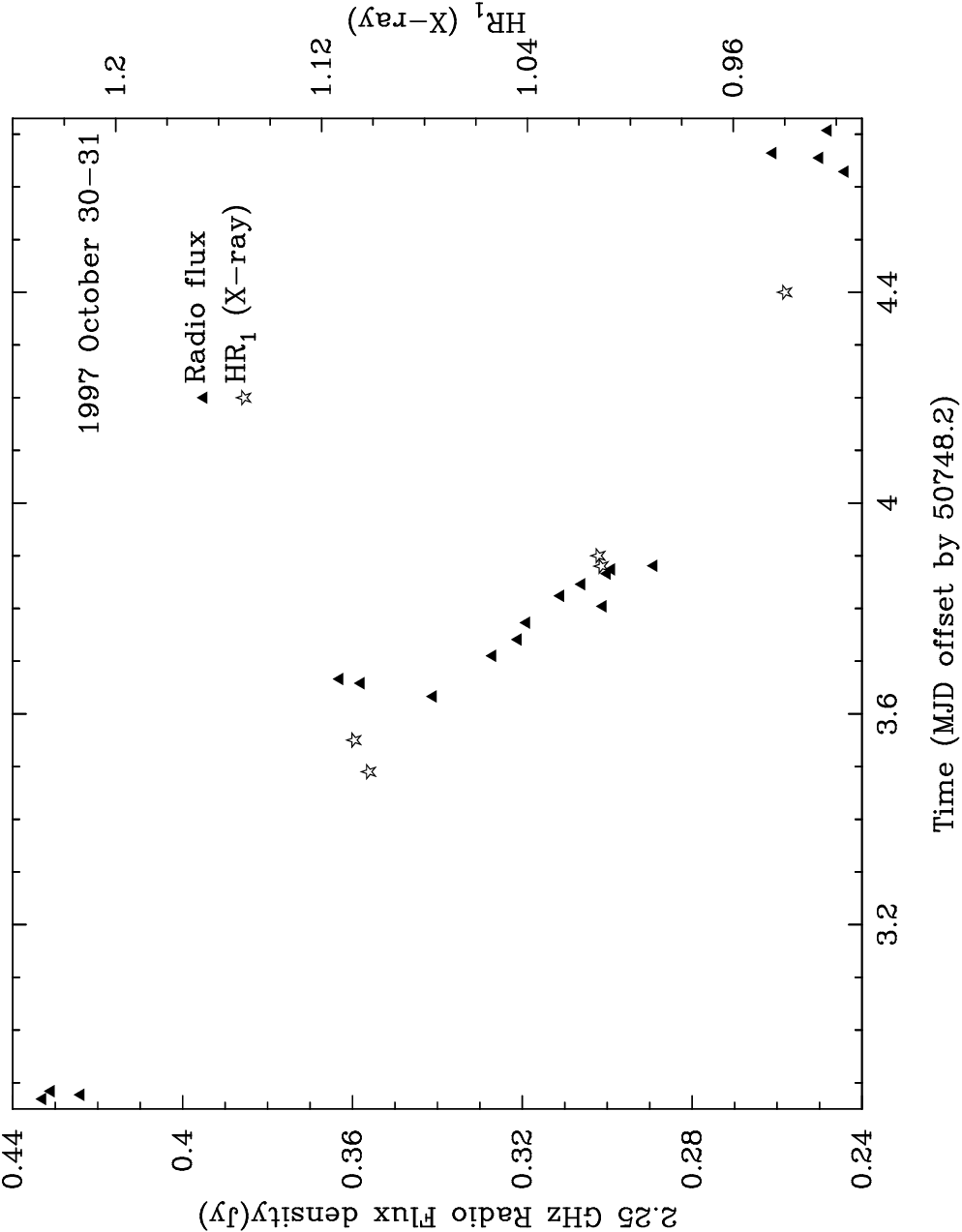}
\figcaption[fig4.ps]{The radio flux at 2.25 GHz from the NSF-NRAO-NASA Green
Bank Interferometer is plotted for 1997 October 30-31 (data for $\sim$ 48 hours)
on the y-axis (left). The X-ray average color HR$_1$ during the quiescent phase 
(state C) is plotted on the y-axis (right) during the same time 
(data for $\sim$ 24 hours). Timing of X-ray observations falls 
between first and second relativistic radio
jets observed by Fender et al. (1999). \label{fig4}}

\end{figure}

\clearpage

\begin{figure}
\psfig{file=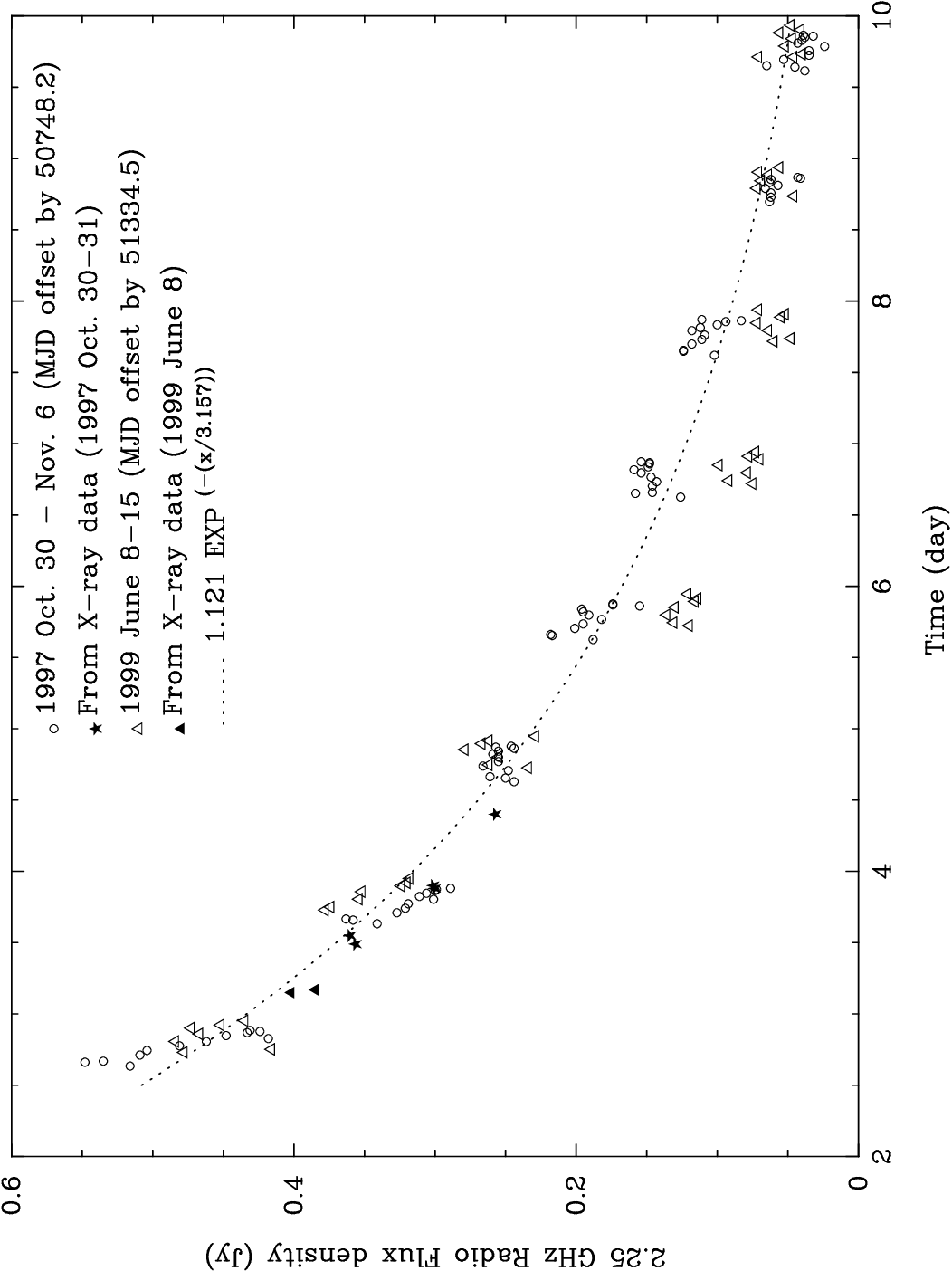}
\figcaption[fig5.ps]{Decay profiles of two relativistic radio flares 
 (class A)  observed on 1997 October 30 - November 6 and on 1999 
June 8-15. The dotted line is an   
exponential fit. The solid points are from X-ray data of both the flares
using the relation $radio_{Jy}=0.615 \times HR_{1}-0.322$ derived from
the data of Figure 4 (for details see text). \label{fig5}}

\end{figure}

\end{document}